\documentstyle[aps,preprint,eqsecnum]{revtex}
\tightenlines
\begin{document}
\newcommand{\be}{\begin{equation}}
\newcommand{\ee}{\end{equation}}
\newcommand{\beq}{\begin{eqnarray}}
\newcommand{\eeq}{\end{eqnarray}}
\newcommand{\ds}{\displaystyle}
\newcommand{\pia}{\mbox{$p_i^{\alpha}$}}
\newcommand{\pjb}{\mbox{$p_j^{\beta}$}}
\newcommand{\la}{\lambda_{\alpha}}
\newcommand{\bla}{\bar{\lambda}_{\alpha}}
\newcommand{\xa}{x^{\alpha}}
\newcommand{\ya}{y^{\alpha}}

\footnotesep .4cm

\vspace{-1in}

\begin{center}
{\bf DYNAMICS OF INDIVIDUAL SPECIALIZATION AND GLOBAL DIVERSIFICATION IN
COMMUNITIES}
\footnote{Journal Ref: {\it Complexity}, Vol. {\bf 3}, No. {3}, 50-56, (1998).}
\vspace{.9cm}

{\bf Vivek S. Borkar}
\footnote{E-mail: {\it borkar@csa.iisc.ernet.in}}\\
{\em Department of Computer Science and Automation, \\
Indian Institute of Science, Bangalore 560 012, India}

\vspace{.6cm}

{\bf Sanjay Jain}
\footnote{Also at Jawaharlal Nehru Centre
for Advanced Scientific Research, Bangalore 560 064, and


\hspace{1cm} Associate Member of ICTP, Trieste. 
E-mail: {\it jain@cts.iisc.ernet.in}}\\
{\em Centre for Theoretical Studies, \\
Indian Institute of Science, Bangalore 560 012, India}

\vspace{.6cm}

{\bf Govindan Rangarajan}
\footnote{E-mail: {\it rangaraj@math.iisc.ernet.in}} \\
{\em Department of Mathematics and Centre for Theoretical Studies, \\
Indian Institute of Science, Bangalore 560 012, India}

\vspace{3cm}

{\bf Abstract}

\vspace{.4cm}

\parbox{6truein}{ We discuss a model of an economic community 
consisting of
$N$ interacting agents. The state of each agent at any time is
characterized, in general, by a mixed strategy profile
drawn from a space of $s$ pure strategies.
The community evolves as agents update their strategy profiles
in response to payoffs received from other agents. The evolution
equation is a generalization of the replicator equation.
We argue that when $N$ is sufficiently large and the payoff matrix
elements satisfy suitable inequalities, the community evolves to retain
the full diversity of available strategies even as individual agents
specialize to pure strategies.}
\end{center}

\newpage
\section{INTRODUCTION}
One of the striking phenomena exhibited by a wide variety of complex adaptive
systems
is that individual agents or components of the system evolve to perform
highly specialized tasks, and at the same time the system as a whole
evolves towards a greater diversity in terms of the kinds of individual
agents or components it contains or the tasks that are performed in it.
Some examples of this include living systems which have evolved increasingly
specialized and diverse kinds of interacting protein molecules, ecologies which
develop diverse species with specialized traits, early
human societies which evolve from a state where everyone shares in a
small number of chores to a state with many more activities performed
largely by specialists, and firms in an economic web that explore and
occupy increasingly specialized and diverse niches.

In this paper we study a mathematical model of economic communities
that exhibits these twin evolutionary phenomena of specialization
and diversity. The system is a community of $N$ (say, human) agents.
There are $s$ strategies or activities each agent can perform
labelled by $i \in S \equiv \{1,2,\ldots,s\}$,
and at the time $t$ the agent $\alpha$ ($\alpha = 1,\ldots ,N$)
performs the activity $i$ with a probability $p^\alpha_i(t)$ (thus
$\sum_{i=1}^s p^\alpha_i(t) = 1 \ \forall \ \alpha,t$). The vector
${\bf p}^\alpha(t) = (p^\alpha_1(t),\ldots,p^\alpha_s(t))$ is called the
mixed strategy profile of agent $\alpha$ at time $t$. If
$p^\alpha_i(t) = \delta_{ij}$ for some $j \in S$
then the agent $\alpha$ is said to pursue the pure strategy $j$
or to have `specialized' in the strategy $j$ at time $t$. The set of
vectors ${\bf p}^\alpha, \ \alpha = 1,\ldots, N$ constitute the basic
degrees of freedom of the model. The dynamics is defined by the
equation
\be
{\dot p}_i^{\alpha}(t) = p_i^{\alpha}(t) \left[\sum_{\beta \neq \alpha} \sum_j
a_{ij}p_j^{\beta}(t) - \sum_{\beta \neq \alpha} \sum_{k,j} p_k^{\alpha}(t)
a_{kj}p_j^{\beta}(t)\right], \ \ \ 1 \leq \alpha \leq N, \ 1 \leq i \leq s,
\label{1.2}\ee
which determines the rate of change of an individual ${\bf p}^\alpha$
in terms of the current mixed strategy profiles of all the agents and
the payoff matrix $A = [[a_{ij}]]$.

We motivate this model as follows: Agents interact with each other on a short
time scale,
receive payoffs based on each other's activity, and update their
individual strategy profiles on a longer time scale so as to increase
their payoffs. As is usual in
game theory, $a_{ij}$ denotes the payoff received by an agent pursuing a pure
strategy $i$ in a single interaction with an agent pursuing the pure
strategy $j$.
Then the average payoff received by the agent $\alpha$ from the rest of
the community
in the period $t$ to $t+\Delta t$ is proportional to
$\Delta t\sum_{\beta \neq \alpha}
\sum_{k,j} p_k^{\alpha}(t) a_{kj}p_j^{\beta}(t)$.
This assumes that every agent interacts equally often with all other
agents and that there is a separation of time scales: $\Delta t$ can
be chosen long enough for there to be a statistically sufficient
number of interactions during the period, yet short enough that
the change in the strategy profiles during this period can be ignored
in the computation of the average payoff.
If $\alpha$ had played the pure strategy $i$ in this period, she would
have received an average payoff proportional to $\Delta t\sum_{\beta \neq
\alpha}
\sum_j a_{ij}p_j^{\beta}(t)$. The agent $\alpha$
increments $p^\alpha_i$ by an amount proportional to $p^\alpha_i$ as well as
to the difference between the average payoff she would have got in this
interval
if she had pursued a pure strategy $i$ and the average payoff she actually
received:
$\Delta p^\alpha_i = c \Delta t p^\alpha_i
[\sum_{\beta \neq \alpha} \sum_j
a_{ij}p_j^{\beta}(t) - \sum_{\beta \neq \alpha} \sum_{k,j} p_k^{\alpha}(t)
a_{kj}p_j^{\beta}(t)]$, where $c$ is a constant.
Equation (\ref{1.2}) follows upon dividing by $\Delta t$, taking the
limit, and rescaling time by a factor $c$. By construction, each
agent makes a positive change in the weight of strategy $i$ in her
own strategy profile if she perceives that the pure strategy $i$ would
give a higher payoff in the current environment than her current strategy
profile, and a negative change if it were to give a lower payoff.

Eq. (\ref{1.2}) is nothing but the `multipopulation replicator equation'
discussed in \cite{Weibull} (and references therein).
There each $\alpha$ represents a population, and
$p^\alpha_i$
represents the fraction of individuals in the population $\alpha$
pursuing the strategy $i$. Since for us each $\alpha$ represents an
individual and not a population, we refer to dynamics specified by
(\ref{1.2})\ as simply the generalized
replicator dynamics (GRD). By contrast the replicator dynamics
(which we hereafter refer to as the `pure replicator dynamics' (PRD))
is given by (see \cite{HS})
\be
{\dot x}_i(t) = x_i(t) [ \sum_j a_{ij}x_j(t) - \sum_{k,j} x_k(t) a_{kj}
x_j(t)], \ \ \ i=1,\ldots,s.
\label{1.1}\ee
This is a standard model in evolutionary biology describing the growth
and decay of $s$ species under selection pressure with $x_i$ representing
the fraction of the $i^{th}$ species in the population.
PRD and its variants are also
extensively studied in economics in game theory as models for dynamical
selection of equilibria (see, e.g., \cite{Mailath}\cite{FL}).
Its generalizations have also been studied in the context of the
emergence of organizations in complex adaptive systems (see \cite{SFM}
and references therein).  For extensive accounts of more recent
contributions to PRD and further references, see the recent books
\cite{LS,FV}.  

We view GRD as a model of learning in a community of $N$ interacting agents.
The agents are identical in that each is capable of pursuing the same
set of strategies with the same payoffs.
This is a non-cooperative game in which the agents
act selfishly (each is concerned with increasing her own payoff without
consideration of impact on others or the community), and
exhibit bounded rationality (no anticipation of others' strategy, merely a
response to the current aggregate behaviour of others).
There is no global organizing agency at work, individual actions alone are
responsible for the evolution of the system.

Nevertheless, we shall argue that the community as a whole seems to
exhibit a kind of global organization under
certain circumstances. Individual agents tend to specialize, while
the community as a whole retains its diversity, i.e., each pure
strategy is pursued by some agent or the other. We attempt to find
conditions on the parameters of the model (the size $N$ of the community
and the $s\times s$ payoff matrix $A$) such that this behaviour
occurs. While most of the time we work with a strategy space of
a fixed size (and refer to diversity as the maintenance of all
strategies in this fixed size space) the results also have bearing
on the conditions under which new strategies can enter the community.

Section 2 sets the notation and
discusses some relationships between PRD and GRD.
In section 3 we identify conditions under which
attractors of GRD can exhibit simultaneously
specialization and diversity, and characterize these attractors quantitatively.
Section 4 summarizes the results,
discusses their possible significance and outlines some open
questions. Due to
constraints on space, proofs for some of the results have not been included
in this paper. These and other generalizations of our results are the
subject of a detailed follow-up paper\cite{preprint}.

\section{RELATIONSHIPS BETWEEN GRD AND PRD; INTERIOR EQUILIBRIA}

\subsection{GRD preliminaries}

\noindent {\bf Notation, definition of specialization and diversity}\hfill
\break
Let $J$ denote the simplex of $s$-dimensional probability vectors:
\be
J=\{ {\bf x} = (x_1,\cdots,x_s)^T \in {\bf R}^s |\sum_{i=1}^s x_i= 1, x_i
\geq 0 \}.
\label{jdef}
\ee
$J$ is the full configuration space of PRD, and is
invariant under it.

The configuration space of GRD will be denoted $J^N = \Pi_{\alpha = 1}^N
J^{(\alpha)}$ where $J^{(\alpha)}$ is a copy of $J$ for
the $\alpha^{th}$ agent. A point of $J^N$ will be denoted
$p = ({\bf p}^1,{\bf p}^2, \ldots ,{\bf p}^N)$, where
${\bf p}^{\alpha} = (p_1^{\alpha},p_2^{\alpha}, \ldots ,p_s^{\alpha}) \in
J^{(\alpha)}$.
$J^N$ is invariant under GRD, as the norm of every ${\bf p}^{\alpha}$
is preserved under (\ref{1.2}).

A point of $J^N$ at which every agent has specialized to some strategy
or the other will be referred to as a corner of $J^N$,
and at such a point we say that the community is `fully specialized'.
It is evident that a corner is an equilibrium point of (\ref{1.2})\ since
$\dot{p}^\alpha_i$ vanishes if $p^\alpha_i$ does and (\ref{1.2})\
preserves norm, hence we often refer to a corner as a `corner equilibrium
point' or CEP. A CEP can be characterized by an
$s$-vector of non-negative integers ${\bf n} = (n_1, \ldots , n_s)$
where $n_i$ is the number of agents pursuing the pure strategy $i$
at the CEP, $1 \leq i \leq s$ (thus $\sum_i n_i = N$).  Two
CEPs with the same associated ${\bf n}$ vector are interchangeable,
since they differ only in the identity of the agents, irrelevant for
our purposes.

The set $F_k \equiv \{p\in J^N|p^\alpha_k=0 \ \forall \ \alpha\}$
for any $k \in S$ is the subset of the boundary of $J^N$ where
all agents have opted out of strategy $k$. At the `face' $F_k$,
strategy $k$ becomes extinct from the population and the full
diversity of strategies is lost. 
The community will be said to exhibit `diversity' at all points
that do not belong to some $F_k$. Note that
we use the word `diversity' not to signify the variation between
individual agents, but to indicate that all strategies
are supported. Indeed we can have no variation but full diversity
if all agents pursue the same mixed strategy: for all $\alpha$,
${\bf p}^{\alpha} = {\bf c} \in J^{\circ}$. (The superscript
$^\circ$ for any set denotes its relative interior.)
When ${\bf p}^{\alpha}$ is
independent of $\alpha$, the community is completely `homogeneous'
since all agents are doing the same thing.
The community can be fully specialized and diversified at the same
time: each agent chooses a pure strategy and every strategy is chosen
by some agent or the other. This corresponds to CEP with ${\bf n}$
such that each $n_i$ is nonzero, which will be called a `fully diversified'
CEP or FDCEP. By contrast, CEP where one or more strategies becomes
extinct (some components of ${\bf n}$ are zero) will be called
non-FDCEP.

In this paper we are primarily interested in studying
the circumstances in which FDCEP are the preferred attractors of
the dynamics, since in that case individual specialization and
global diversity will arise dynamically in the community.

\vskip 0.5cm
\noindent{\bf Differences between PRD and GRD}\hfill\break
If the initial point of a trajectory in GRD is homogeneous, the trajectory
remains homogeneous for all time, and evolves according to (\ref{1.1})\
except that the time is speeded up by a factor of $N-1$.
The sum $\bar {x}_i \equiv (1/N) \sum_{\alpha =1}^N p_i^{\alpha}$
equals the probability that strategy $i$ is being played in the
entire community, and is therefore the analogue of $x_i$ in PRD.
We can ask how $\bar{x}_i$ evolves in GRD. It is easy to see that
\be
\dot{\bar{x}}_i = N[\bar{x}_i \sum_j a_{ij} \bar{x}_j
- \sum_{k,j} x_{ik} a_{kj} \bar{x}_j
- {1 \over N} \sum_j x_{ij} a_{ij}
+ {1 \over N} \sum_{k,j} x_{ikj} a_{kj}],\label{barx}
\ee
where $x_{ik} \equiv (1/N)\sum_{\alpha} p^{\alpha}_i p^{\alpha}_k$ and
$x_{ikj} \equiv (1/N)\sum_{\alpha} p^{\alpha}_i p^{\alpha}_k p^{\alpha}_j$.
The r.h.s. of (\ref{barx})\ is not proportional
to the r.h.s. of (\ref{1.1}), except for homogeneous trajectories
in which case $x_{ik} = \bar{x}_i \bar{x}_k$,
$x_{ikj} = \bar{x}_i \bar{x}_k \bar{x}_j$. Thus in general $\bar{x}_i$
does not follow the PRD. One might have hoped that when the
number of agents $N$ is large $\bar{x}_i$ follows PRD, but even
that is not the case due to variation among the agents. For example,
at the corner ${\bf n}$, the difference between
$x_{ik}$ and $\bar{x}_i \bar{x}_k$ is $n_i(N \delta_{ij} - n_j)/N^2$,
which is comparable to the former two even for large $N$
(except for homogeneous corners).

One of our results in this paper is that
even though variation among agents, which is generic in GRD,
causes the evolution of $\bar{x}_i$ to
be different from PRD, under suitable conditions
$\bar{x}_i$ nevertheless converges to the interior equilibrium point of
PRD.

\subsection{The interior equilibria of GRD}

Consider an interior equilibrium point (IEP) $p$ of (\ref{1.2}).
By definition no $p^\alpha_i$ is zero in the interior of $J^N$. Therefore
the bracket [ ] on the r.h.s. of (\ref{1.2})\ must vanish for all $\alpha,i$.
Define $x_0^{\alpha} \equiv \sum_{\beta \neq \alpha}
\sum_{i,j} p_i^{\alpha} a_{ij} p_j^{\beta}$, and
$v^{\alpha}_i \equiv \sum_{\beta \neq \alpha} p^{\beta}_i$.
Then $\sum_{i=1}^s v^{\alpha}_i = N-1 \ \forall \ \alpha$, and
the interior equilibrium condition can be written as
\be
B {\bf X}^{\alpha} = (N-1){\bf E}_0 \ \ \ \forall  \ \ \ \alpha,
\label{GIEP}\ee
where ${\bf X}^{\alpha} \equiv (x^{\alpha}_0, v^{\alpha}_1, v^{\alpha}_2,
\ldots , v^{\alpha}_s)^T$, $B$ is the $s+1$-dimensional matrix
\be
B =
\left( \begin{array}{ccccc}
0&1&1&\cdots&1 \\
-1&&&&         \\
-1&&A&&        \\
\vdots &&&&    \\
-1&&&&
\end{array} \right), \label{B}
\ee
and ${\bf E_0}$ is the $s+1$-dimensional unit vector $(1,0,0,\cdots,0)^T$.

It is not difficult to see (details in \cite{preprint})
that (\ref{GIEP}) has an isolated solution if and only
if the following condition holds:
\begin{description}
\item{} {\bf A1}: \ \ \ $u_i \neq 0 \ \ \forall \ \ i$, and all $u_i$ have the
same sign, where $u_i$ denotes the co-factor of $B_{0i}$.
\end{description}
For the sake of notational simplicity, we have denoted 
the cofactor of $B_{0i}$ by $u_i$ instead of $u_{0i}$ thereby 
suppressing the fixed first index.
Under the above condition ({\bf A1}) $\det B = \sum_{i=1}^s u_i \neq 0$, and
the solution is unique and given by $v^{\alpha}_i = (N-1)x_i \
\forall \ \alpha$. Here,
\be
x_i = u_i/{\rm det} B
\label{IEP}\ee
is nothing but the
$i^{th}$ coordinate of the unique isolated interior equilibrium point
of PRD. (Note that {\bf A1} is also
the necessary and sufficient condition for PRD to have an
isolated IEP, which, if it exists, is unique.)
Since $p^{\alpha}_i - p^{\beta}_i =
v^{\beta}_i - v^{\alpha}_i = 0$, it follows that the equilibrium
point is homogeneous and given by $p^{\alpha}_i = x_i$. Thus we
have proved \hfill\break
{\bf Theorem 2.1} There exists at most one isolated equilibrium in
the interior of $J^N$. It exists if and only if {\bf A1} is satisfied
and then it is homogeneous (all agents pursue the same mixed strategy),
and coincides with the isolated interior equilibrium point of PRD,
$p^{\alpha}_i = x_i \ \forall \ \alpha,i$.

\section{CORNER EQUILIBRIA OF GRD: DIVERSIFICATION WITH SPECIALIZATION}

\subsection{Stability of corner equilibria}

The IEP of GRD is always unstable to small perturbations. This is a
consequence of the following theorem proved in \cite{Weibull}:

{\bf Theorem 3.1} An equilibrium point of (\ref{1.2})
is asymptotically stable if and only
if it is a strict Nash equilibrium.  Further, any compact set in the
relative interior of a face cannot be asymptotically stable.

Note that strict Nash equilibria are perforce pure strategy Nash
equilibria and therefore correspond to CEP.
As a consequence of this theorem, a trajectory either
eternally moves around in the relative interior of
some face or the interior of $J^N$ coming arbitrarily close to its
boundaries and corners (the case of non-compact attractor),
or it converges to a corner of $J^N$.
It is possible to construct payoff matrices
for which there are no asymptotically stable corners in $J^N$, whereupon
the former situation obtains. However, our numerical work
with $3 \times 3$ payoff matrices suggests that this happens rarely
(i.e., in a relatively small region of ${\bf R}^{3 \times 3}$);
for most payoff matrices asymptotically stable corners do exist
for most values of $N$.  
Further, we randomly generated ten
$3 \times 3$ payoff matrices and numerically integrated
the GRD equations for long times for each payoff matrix with ten
randomly chosen initial conditions. When this was done with $N=5$,
in 90 out of the 100 cases the dynamics converged to a corner. 
With $N=10$, all 100 cases converged to a corner. This suggests
that typically, at large $N$, not only do asymptotically
stable corners exist, but also that their
basins of attraction cover most of $J^N$.
Thus corners seem to be the most common attractors in GRD. These
are numerical indications and need to be made more precise.
In our interpretation of the model, a corner
corresponds to a fully specialized community. The above theorem and
numerical evidence therefore suggest
that specialization of all the agents is the most common outcome in GRD.

At the CEP ${\bf n}$, the
payoff to an agent playing the $j^{th}$ pure strategy from the other
$N-1$ agents is
\be
{\cal{P}}_j = \sum_{k \neq j}^s a_{jk} n_k + (n_j - 1)a_{jj}
=  \sum_{k=1}^s a_{jk} n_k - a_{jj} = P_j - a_{jj}, \label{payoff}
\ee
where $P_j \equiv \sum_{k=1}^s a_{jk} n_k$. If this agent were to suddenly
switch to the $i^{th}$ pure strategy ($i \neq j$), all other agents
remaining at their respective pure strategies, then for this agent the
payoff would change to
$\sum_{k \neq j}^s a_{ik} n_k - a_{ij}(n_j - 1) = P_i - a_{ij}$.
Thus the increase in payoff for an agent playing the $j^{th}$ pure strategy
at the FDCEP ${\bf n}$ (and this assumes $n_j \neq 0$)
in switching to the $i^{th}$ pure strategy is
\be
\lambda_{ij} = P_i - P_j - h_{ij}, \quad \quad \quad
h_{ij} \equiv a_{ij} - a_{jj}. \label{evs}
\ee
Therefore, ${\bf n}$ is a strict Nash equilibrium if for every $j$ such that
$n_j \neq 0$, the conditions
\be
\lambda_{ij} < 0 \label{Nash}
\ee
are satisfied for all $i\neq j$.
At a FDCEP, all $n_j$ are nonzero and this is a set of $s(s-1)$ conditions.
At a non-FDCEP where only $s' < s$ components of ${\bf n}$ are nonzero,
the number of conditions is smaller, $s'(s-1)$.

{}From Theorem 3.1, these are identical to the conditions
for the asymptotic stability of the FDCEP's associated with ${\bf n}$.
In fact, one can show that $\lambda_{ij}$ given by (\ref{evs})\
are precisely the eigenvalues of the Jacobian matrix of (\ref{1.2})\
linearized around a corner of $J^N$ characterized by ${\bf n}$.

\subsection{Stability of fully diversified corners}

{\bf Theorem 3.2}:
Let PRD admit an isolated IEP ${\bf x}$. That is,
condition {\bf A1} holds [cf. Section II]. Let ${\bf n},{\bf n'}$
be any pair of asymptotically stable
FDCEP's of GRD with $N \geq s$. Then
\begin{description}
\item{({\it i}) all
components of the difference ${\bf n'} - {\bf n}$ are bounded by a function
of $A$ alone, not of $N$, and}
\item{({\it ii}) $\lim_{N \rightarrow \infty} {n_i \over N} = x_i$.}
\end{description}

The proof is given in Appendix A.  

The significance of this theorem is that it characterizes the FDCEP
that are attractors of the dynamics.
If the community is going to end up in a fully specialized and
diversified configuration, the theorem quantifies the relative
weights of all strategies that will obtain in that configuration:
these relative weights are forced to be `close' to the IEP configuration
given by (\ref{IEP}).
The theorem does not guarantee the existence of a stable FDCEP.
One can prove the existence of an infinite set of values of $N$
at which stable FDCEP are guaranteed to exist under the conditions of
the theorem. One can also identify sufficient conditions for
the existence of stable FDCEP for any $N \geq s$. These
are presented in \cite{preprint}.

\subsection{Instability of non-fully diversified corners}

We would like to define GRD as possessing diversity if all trajectories
in the faces $F_k$ become unstable at some time or the other
with respect to perturbations that take them away from these faces. With
this in mind we
now study corners at which one or more strategies become extinct and 
determine the conditions under which all such corners
become unstable. Then under small perturbations the population will
dynamically flow out of such corners, eliminating specialized configurations
that do not carry the full diversity of strategies. As remarked
earlier, the number of conditions to be satisfied by a non-FDCEP
to be stable is less than the number to be satisfied by a FDCEP.
Thus a priori, things seem to be loaded against diversification.
As we shall see,
some further structure will need to be imposed on $A$ in order to
make the non-FDCEP unstable. At this point we do not have the general
conditions for arbitrary $s$, but some insight gleaned from 
special cases $s=2,3$.

\noindent {\bf ${\bf s=2}$, ${\bf N}$ arbitrary}\hfill\break
In this case conditions (\ref{Nash})\ can be studied exhaustively.
There are generically four cases.
\begin{description}
\item{Case 1:} $a_{11} > a_{21}$ and $a_{22} > a_{12}$: Both $(N,0)$ and
$(0,N)$ are asymptotically stable, other corners are not.

\item{Case 2:} $a_{11} > a_{21}$ and $a_{22} < a_{12}$: $(N,0)$ is the only
asymptotically stable corner.

\item{Case 3:} $a_{11} < a_{21}$ and $a_{22} > a_{12}$: $(0,N)$ is the only
asymptotically stable corner.

\item{Case 4:} $a_{11} < a_{21}$ and $a_{22} < a_{12}$:
The only asymptotically stable corners $(n_1,n_2)$ (with $n_1 + n_2 =N$)
are those for which $n_1 \neq 0$, $n_2 \neq 0$, and
furthermore,
$\frac{(n_2-1)}{n_1} h_{12} < h_{21} < \frac{n_2}{(n_1-1)}h_{12})$
if $n_2 < N-1$, and
$\frac{(n_2-1)}{n_1} h_{12} < h_{21}$
if $n_2 = N-1$.
\end{description}

Cases 2 and 3 correspond to dominated strategies.  (The cases with one or
more equalities instead of inequalities have been disregarded as nongeneric.
In any case, they are not difficult to handle.)  The case of interest to
us is the last one, which shows diversification. It is convenient to
introduce the
\begin{description}\item[]
{\bf Definition}: $A$ is diagonally subdominant if $a_{ii} < a_{ji} \ \forall
\ j \neq i, \ \{i,j\} \subset S$.
\end{description}
That is, $h_{ij} > 0 \ \forall \ i \neq j$.
{}From the above exhaustive list it follows that the
condition
\begin{description}
\item{\bf A2}: \  \ \  $A$ is diagonally subdominant,
\end{description}
is the necessary and sufficient condition for non-FDCEP to be
unstable (for generic $A$). If {\bf A2} is satisfied, the only
asymptotically stable CEP are the FDCEP, for which Theorem 3.2
applies. (Note that for $s=2$, {\bf A2} implies {\bf A1}, the
IEP is given by
${\bf p}^{\alpha} = {1 \over {h_{12} + h_{21}}}(h_{12},h_{21})$ for
all $\alpha$, and the inequalities involving ${\bf n}$ in Case 4 above are
equivalent to
the statement that $(1/N)(n_1,n_2)$ must be close to this IEP
for arbitrary $N$ and converge to it as $N\rightarrow \infty$.)

Note that GRD remains invariant under addition of an arbitrary
constant to any column of the payoff matrix.  Thus we may replace 
$a_{ij}$ by $h_{ij}$, thus obtaining a matrix which under {\bf A2}
has zero diagonal elements and nonnegative off-diagonal elements.
It is interesting that these conditions also arise 
in PRD in the context of population genetics and ecological models \cite{GNG}\ 
as well as in models of catalytic networks of 
chemically reacting molecules \cite{HSSW}. 

\vskip 0.3cm \noindent
{\bf ${\bf s=3}$, ${\bf N}$ arbitrary}\hfill\break
For $s=3$, {\bf A2} no longer implies {\bf A1}; the latter is an
independent condition. We now state
{\bf Theorem 3.3}
For $s=3$, if both {\bf A1} and {\bf A2} hold, then there exists
a positive number $N_0$ depending on $A$ such that for all
$N > N_0$, all non-FDCEP are unstable.

The proof of this theorem can be found in Appendix B.

We remark that while $N_0$ is finite, it may, depending upon
$A$, be much larger than three.
The above result can be further generalized (with the imposition of an
additional condition) to prove that for $s=3$ all points in
$F_k$ are unstable for sufficiently large $N$ \cite{preprint}. Note that
our notion of diversity for GRD is related to the notions
of `permanence', `persistence', etc. introduced for PRD (see \cite{HS}\
and references therein).
PRD is said to exhibit permanence if every interior solution has
components that remain bounded away from zero by a common constant
$\delta > 0$.  Strong persistence, in turn, is the weaker requirement
where $\delta$ is trajectory dependent and persistence the even weaker
requirement that each component of an interior trajectory not
converge to zero.  The biological implications are obvious: the 
concept is clearly related to survival of species.  The corresponding
phenomenon here is the survival of policies.  The
conditions for, e.g., permanence in PRD (see \cite{HS}) may quite generally 
play a role in discussions of diversity in GRD.

\section{DISCUSSION AND CONCLUSIONS}

To summarize:

(i) We have considered the equation (\ref{1.2}) as
a model of evolution of a community of $N$ agents, each agent being capable of
performing any mix of a set of $s$ strategies, and modifying
her mix depending upon the payoff received from other agents.
We have studied some properties of the attractors of this system
to gain insight on how the community is expected to evolve.

(ii)
This model can exhibit specialization of the agents into pure
strategies. Evidence for this comes from the previously known
Theorem 3.1, supported with our numerical observations. While individual
specialization seems to be the most common outcome in this model, it would
be interesting to characterize more precisely the circumstances
in which specialization is guaranteed, i.e., when corners are the
only attractors, and when not.

(iii) We have shown that under
suitable conditions, while each agent specializes to a single
pure strategy, it is guaranteed that the community as a whole
preserves the full diversity of strategies. These
are that the community be sufficiently large ($N$ should be
larger than a number $N_0$ that depends upon the payoff matrix),
and the payoff matrix itself should satisfy {\bf A1}
(existence of an isolated interior equilibrium point) and {\bf A2} (diagonal
entries of $A$ be smaller than other entries in the same column).
These guarantee (for upto three strategies) that all corners where
one or more strategy becomes extinct are unstable to small perturbations
(Theorem 3.3). To identify sufficient conditions for larger $s$
(and necessary and sufficient conditions for $s \geq 3$) is
a task for the future. The appearance of a lower limit on the size
of the community in this context (which could be much
larger than the number of strategies) is interesting.

(iv) Within the set of configurations where the community would
exhibit full specialization and diversity (the FDCEP), we
have given a quantitative criterion as to which ones will be
the attractors (Theorem 3.2). $n_i/N$
(where $n_i$ is the number of agents pursuing the pure strategy
$i$ at the attractor) is forced to be close to $x_i$
and equal to it in the large $N$ limit, where $x_i$ is given by (\ref{IEP})
and is the relative weight of the $i^{th}$ strategy at the interior
equilibrium point of PRD.
This constraint is a consequence of the fine balance that exists
for every agent at a strict Nash equilibrium; any strategy switch
for any agent reduces her payoff. This fine tuning, caused by the
interaction of the agent with other agents, is a kind of organization
exhibited by the system.

The conditions for the instability of non-FDCEP
(Theorem 3.3) may also be relevant to
the question: when does a society accept an innovation? For consider
a community of a large number of agents but with only two strategies,
$1$ and $2$, at a stable corner where $n_1$ agents pursue the
pure strategy $1$ and $n_2 = N-n_1$
agents the pure strategy $2$ (neither $n_1$ nor $n_2$ is zero).
Since this corner is assumed stable,
the $2 \times 2$ matrix $A$ satisfies condition {\bf A2}
(diagonal subdominance). Now imagine that a new strategy $3$ arises
thereby enlarging the payoff matrix to a $3 \times 3$ matrix $A'$
containing $A$ as a $2 \times 2$ block. In the new context
the earlier state of the community will be described by a three
vector ${\bf n}= (n_1,n_2,0)$, which is in the face $F_3$.
Now if the new payoff matrix satisfies {\bf A1,A2}, and $N$
is sufficiently large, then from Theorem 3.3, this configuration
is unstable with respect to perturbations in which one of the
agents begins to explore the new strategy.
Thus if this agent were to explore the new strategy  
ever so slightly, her payoff would increase and a small 
perturbation of the community would grow until it settles down 
in another attractor.
The new attractor if described by Theorem 3.2 would have the
property that
a finite fraction of the population pursues the new strategy:
the innovation has been accepted by the society. Thus the
conditions {\bf A1, A2} of Theorem 3.3 indicate what the payoffs of a new
strategy (innovation) should be with respect to the existing ones, if the
innovation is to be guaranteed acceptance. (Conditions that are
both necessary and sufficient for diversity would considerably strengthen
the above remarks.)

It is worth mentioning that conditions {\bf A1, A2} are not
equalities but inequalities. Thus there is no fine tuning of
parameters needed; the behaviour discussed above emerges whenever
parameters cross certain thresholds.

It may be interesting to consider the `economic significance'
of conditions which play an important role
in preserving the full diversity of strategies. For example, diagonal
subdominance, when translated as `each pure strategy gives more
payoff to other pure strategies than to itself',
carries a shade of an `altruism' of sorts (at the level of
strategies, not individuals). Note that PRD with
a payoff matrix in which diagonal entries are zero and offdiagonal ones
greater than or equal to zero is called a `catalytic network'
\cite{HS}. The general message might be that if the initial set of allowed
strategies is chosen with the `right vision' (read `right payoffs'), then,
even a community of identical and selfish individuals, if large enough, will
exhibit diversity and accept only the `right' innovations.

\section*{APPENDIX A}

The proof of Theorem 3.2 in Section IIIB follows:

{\bf Proof}:
Note that $P_i - P_j$ figures in both $\lambda_{ij}$
and $\lambda_{ji}$. Therefore the $s(s-1)$ conditions (\ref{Nash}) can
be written in terms of $s(s-1)/2$ ``double-sided'' inequalities
\be
-h_{ji} < P_i - P_j < h_{ij}. \label{stab}
\ee
Define $z_i \equiv P_i - P_{i+1}$ for $i = 1,\ldots ,s$, with $P_{s+1}
\equiv P_1$. Then $z_i = \sum_{j=1}^s c_{ij} n_j$ with
$c_{ij} \equiv a_{ij} - a_{i+1,j}$,
where it is again understood that $a_{s+1,j} \equiv a_{1j}$. Now,
since all the $n_j$ are not independent, let us express $z_i$ in terms
of only $n_1,\ldots ,n_{s-1}$ by eliminating
$n_s = N - (n_1 + \cdots + n_{s-1})$. This gives $z_i = y_i + c_{is} N$
where $y_i \equiv \sum_{j=1}^{s-1} d_{ij} n_j$ and
$d_{ij} \equiv c_{ij} - c_{is}$, $i,j = 1,\ldots ,s-1$.
With this notation, consider the subset of $s-1$ inequalities
obtained by setting $j=i+1$ in (\ref{stab}), with $i = 1,\ldots ,s-1$.
These involve $z_i$ and take the form
\be
-h_{i+1,i} - c_{is} N < y_i < h_{i,i+1} - c_{is} N, \quad i=1,\ldots ,s-1.
\label{stabi}
\ee
These inequalities mean that for any stable FDCEP ${\bf n}$, the
$y_i$, which are linear
combinations of $n_1,\ldots ,n_{s-1}$, are constrained to be in an
open interval of the real line. While the location of this interval
is $N$ dependent, it follows from (\ref{stabi}) that the size
of this interval is finite, independent of $N$, and depends only on the
payoff matrix (for $y_i$ the size of the interval is
$h_{i+1,i} + h_{i,i+1}$).

If the $s-1$ dimensional matrix $D = (d_{ij})$ has an inverse, we can
invert $y_i \equiv \sum_{j=1}^{s-1} d_{ij} n_j$ to express the
$n_j$ in terms of $y_i$. Then,
(\ref{stabi}) will get converted into inequalities for
$n_1,\ldots ,n_{s-1}$. Since the vector
$\tilde{{\bf y}} = (y_1,\ldots ,y_{s-1})$ in the $s-1$ dimensional cartesian
space whose axes are the $y_i$ is constrained by (\ref{stabi}) to lie
in a (rectangular) parallelepiped, the vector
$\tilde{{\bf n}} = (n_1,\ldots ,n_{s-1})$ in the $s-1$ dimensional cartesian
space whose axes are the $n_i$ will also lie in a (in general oblique)
parallelepiped which is the image, under $D^{-1}$, of the rectangular
parallelepiped in $y$-space defined by (\ref{stabi}). Again, while
the location of the parallelepiped in $n_1,\ldots ,n_{s-1}$ space will
depend upon $N$, its size, i.e., its extent along any of the coordinate axes,
will be independent of $N$. This is because
the matrix $D^{-1}$, if it exists, depends only on the payoff matrix
and not on $N$. Therefore, if $D^{-1}$ exists, the differences
in $n_i$, $i=1,\ldots ,s-1$ for all FDCEP are bounded by some function
of $A$ alone, not of $N$. The same is true for $n_s$ also since
$\sum_{i=1}^s n_i = N$. One can show \cite{preprint} that
$\det D = \det B$.
The existence of $D^{-1}$ is thus guaranteed by condition
{\bf A1}, completing the
proof of the first part of Theorem 3.2.

To prove the second part of Theorem 3.2, divide all sides of
(\ref{stab}) by $N$ and take the limit $N \rightarrow \infty$.
This yields $\lim_{N \rightarrow \infty} [{P_i\over N} - {P_j \over N}] = 0$.
Defining $x'_i \equiv \lim_{N \rightarrow \infty} (n_i/N)$
along an appropriate subsequence independent of $i$, this
is equivalent to the statement that $\sum_{k=1}^s a_{ik} x'_k$ is
independent of $i$, which implies that ${\bf x'}$ is the same
as ${\bf x}$, the IEP of PRD.  $\Box$

\section*{APPENDIX B}

The proof of Theorem 3.3 in Section IIIC follows:

{\bf Proof}:
Any non-FDCEP ${\bf n}$ must belong to some $F_k$, in this case to
$F_1$, $F_2$ or $F_3$. For every $i$ such that the component $n_i$
of ${\bf n} \in F_k$ is nonzero, consider the eigenvalue
\be
\lambda_{ki} = P_k - P_i - h_{ki}.\label{lac}
\ee
{}From the discussion of Eq. (\ref{evs})\ it follows that if any one
(or the largest) of the $\lambda_{ki}$ at ${\bf n}$ is greater than
zero, then the CEP ${\bf n}$ is unstable against perturbations
in which an agent pursuing the pure strategy $i$ moves towards
strategy $k$ (i.e., the perturbations which restore the
extinct strategy $k$ will then grow).

For concreteness consider $F_3$. Corners of $F_3$ are of two types.
\hfill\break \noindent Case 1: Only one strategy survives at the
corner. Then ${\bf n} = (N,0,0)$ or $(0,N,0)$. In the former case
(\ref{lac})\ implies $\lambda_{31} = (N-1)h_{31}$ and in the latter
case $\lambda_{32} = (N-1) h_{32}$. By {\bf A2} both corners are unstable.
\hfill\break
\noindent Case 2: Both strategies $1$ and $2$ survive at the corner of
$F_3$. Then ${\bf n} = (n_1,n_2,0)$ with both $n_1$ and $n_2$
positive integers and $n_1 + n_2 = N$. There are then two eigenvalues from
(\ref{lac}),
$\lambda_{31} = h_{31} n_1 + h_{32} n_2 - h_{12} n_2 - h_{31}$, and
$\lambda_{32} = h_{31} n_1 + h_{32} n_2 - h_{21} n_1 - h_{32}$. Let us
assume that this corner is stable, hence both $\lambda_{31},\lambda_{32}$ are
negative. The condition $\lambda_{31} < 0$ (upon eliminating
$n_1 = N- n_2$) reduces to
$(h_{12} + h_{31} - h_{32})n_2 > (N-1) h_{31}$. Since $n_2, N-1,
h_{31}$ are all positive this means that the combination
$h_{12} + h_{31} - h_{32}$ is also positive, and
\be
{{(N-1) h_{31}} \over {h_{12} + h_{31} - h_{32}}} < n_2.
\ee
Similarly $\lambda_{32} < 0$ implies that $h_{21} + h_{32} - h_{31}$ is
positive (as can be seen by eliminating $n_2$) and further,
\be
n_2 < {(N-1) {(h_{21}-h_{31})} \over {h_{21} + h_{32} - h_{31}}} + 1.
\ee
Combining the two, we get
\be
{{(N-1) h_{31}} \over {h_{12} + h_{31} - h_{32}}} <
{(N-1) {(h_{21}-h_{31})} \over {h_{21} + h_{32} - h_{31}}} + 1,
\ee
which can be rearranged into the form
\be
(N-1)[-h_{21}h_{12} + h_{21}h_{32} + h_{31}h_{12}] <
(h_{12} + h_{31} - h_{32})(h_{21} + h_{32} - h_{31}).
\ee
But the quantity in [ ] on the l.h.s. of this inequality is just $u_3$
(as evaluated from the definition given in {\bf A1}), which is
positive. (The positivity of $\det B$ and hence $u_1,u_2,u_3$ also follows
from {\bf A1} and {\bf A2}.) Thus we have
\be
N < {{(h_{12} + h_{31} - h_{32})(h_{21} + h_{32} - h_{31})}\over u_3} + 1.
\ee
Note that the r.h.s. is a function of $A$ alone and is finite, say $N_0(A)$.
If $N$ is chosen larger than $N_0(A)$, this inequality is
violated. That is, for $N > N_0(A)$, the corner of $F_3$ under consideration
cannot be stable. We have thus proved that under {\bf A1,A2}, all corners
of $F_3$ are unstable for $N > N_0(A)$. Similarly one may consider
$F_1, F_2$, which will yield the same result but with different finite
bounds in place of $N_0(A)$. We can henceforth use $N_0$ for the
largest of the three. The claim follows. $\Box$

\end{document}